\documentclass[pra,twocolumn,aps,showpacs,superscriptaddress]{revtex4-1}
\usepackage{epsfig}
\usepackage{subfigure}
\usepackage{bm}
\usepackage{amsmath}
\usepackage{hyperref}

\begin{document}
\title{Condensed phase of Bose-Fermi mixtures with a pairing interaction}
\author{Andrea Guidini}
\affiliation{School of Science and Technology, Physics Division, 
University of Camerino, Via Madonna delle Carceri 9, I-62032 Camerino, Italy}
\author{Gianluca Bertaina}
\affiliation{Dipartimento di Fisica, Universit\`a degli Studi di Milano, Via Celoria 16, I-20133 Milano, Italy}
\author{Davide Emilio Galli}
\affiliation{Dipartimento di Fisica, Universit\`a degli Studi di Milano, Via Celoria 16, I-20133 Milano, Italy}
\author{Pierbiagio Pieri}
\affiliation{School of Science and Technology, Physics Division, 
University of Camerino, Via Madonna delle Carceri 9, I-62032 Camerino, Italy}
%\affiliation{INFN, Sezione di Perugia, Perugia, Italy}
\date{\today}

\begin{abstract}
We study the condensed phase of a Bose-Fermi mixture with a tunable pairing interaction between bosons and fermions with many-body diagrammatic methods and fixed-node diffusion Quantum Monte Carlo simulations.  A universal behavior of the condensate fraction and bosonic momentum distribution with respect to the boson concentration is found to hold in an extended range of boson-fermion couplings and concentrations.  For vanishing boson density,  we prove that the bosonic condensate fraction reduces to the quasiparticle weight Z of the Fermi polaron studied in the context of polarized Fermi gases, unifying in this way two apparently unrelated quantities.
 \end{abstract}

\maketitle

Bose-Fermi (BF) mixtures with a tunable pairing interaction between bosons and fermions have been actively investigated  in the context of ultra-cold gases \cite{Pow05,Dic05,Sto05,Avd06,Pol06,Roethel07,Bar08,Pol08,Bor08,Mar08,Wat08,Fra10,Yu11,Lud11,Song11,Fra12,Yam12,And12,Ber13,Fra13,Sog13,Gui14},
where the tunability of the BF interaction has been demonstrated and exploited in several experiments  \cite{Osp06,Osp06b,Zir08,Ni08,Wu11,Wu12,Park12,Heo12,Cum13,Bloom13}.   
Previous work has shown that, even at zero temperature, a sufficiently strong BF attraction suppresses completely the boson condensate in mixtures where the number of bosons does not exceed the number of fermions~\cite{Pow05,Fra10,Lud11}. This is due to pairing of bosons with fermions  into molecules, which competes with condensation in momentum space.  In particular, a  first-order phase transition from a superfluid phase with a bosonic condensate, to a normal (molecular) phase without a condensate was recently demonstrated with fixed-node Diffusion Monte Carlo (FNDMC) simulations~\cite{Ber13}.
% In the molecular phase, an interesting depletion effect on the bosonic momentum distribution at low momenta was found in \cite{Fra12} within a T-matrix diagrammatic formalism and confirmed also with FNDMC calculations in \cite{Gui14}. 
 
 Here, we focus on the superfluid phase at zero temperature and present a many-body diagrammatic formalism able to describe this phase from weak to strong BF coupling. Our approach is validated by comparing it with previous~\cite{Ber13} and new dedicated FNDMC calculations. 
By using both methods, we then analyze the condensate fraction and the momentum distributions, and establish a remarkable connection with the polaron problem in polarized Fermi gases.

 {\em Model and diagrammatic formalism} -
The system of our interest is a mixture of bosons of mass $m_{\rm B}$ and number density $n_{\rm B}$, interacting with spinless fermions of mass $m_{\rm F}$ and number density $n_{\rm F}$. The system is dilute, such that the range of all interactions  can be considered smaller than the relevant inter-particle distances. 
The BF pairing interaction can be described then by an attractive contact potential, whose strength is parametrized in terms of the BF scattering length $a_{\rm BF}$ with the same regularization procedure commonly used for Fermi gases~\cite{Sad93,Pie00}.  The interaction between bosons is instead assumed to be repulsive, with scattering length $a_{\rm BB}$ of the order of the interaction range. No interaction between fermions is considered, since short-range interactions are suppressed by Pauli principle. 
We are interested in systems with concentration of bosons $x=n_{\rm B}/n_{\rm F}\le 1$, where a full competition between pairing and condensation is allowed. A natural (inverse) length scale is then provided by the Fermi wave vector $k_{\rm F}\equiv (6\pi^2 n_{\rm F})^{1/3}$, which can be combined with $a_{\rm BF}$  to define the dimensionless coupling strength $(k_{\rm F} a_{\rm BF})^{-1}$. For weak attraction $a_{\rm BF}$ is small and negative, such that $(k_{\rm F} a_{\rm BF})^{-1}\ll -1$ and perturbation theory is applicable~\cite{Alb02,Viv02}. For strong attraction $a_{\rm BF}$ is small and positive,  such that $(k_{\rm F} a_{\rm BF})^{-1}\gg 1$, and the system becomes effectively a mixture of molecules and unpaired fermions (if any), which can be described again by perturbation theory  (now for a Fermi-Fermi mixture). The most challenging regime is then the intermediate one, where $|k_{\rm F} a_{\rm BF}|\gtrsim1$ and perturbation theory fails. 
\begin{figure}[t]
\begin{center}
\includegraphics[width=8.5cm]{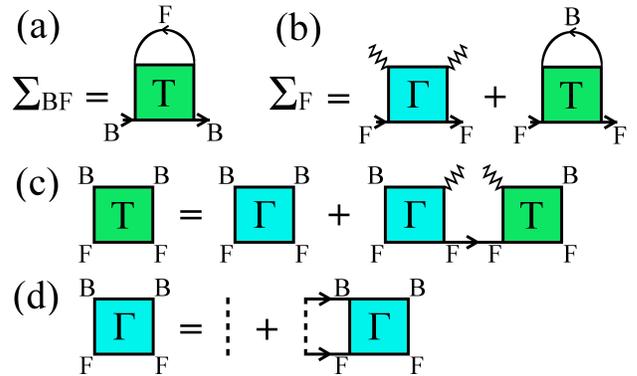}
 \caption{Feynman's diagrams for  $\Sigma_{\rm BF}$, $\Sigma_{\rm F}$,  T, and  $\Gamma$ . Full lines correspond to bare boson (B) and fermion (F) Green's functions, dashed lines to bare BF interactions, zig-zag lines to condensate factors $\sqrt{n_0}$.} 
\label{diagrams}
\end{center}
\end{figure}

Previous experience with the similar problem of the BCS-BEC crossover \cite{Per04,Pie05,Pal10,Pal12} suggests that  selection of an appropriate class of diagrams might provide a reliable approach even in this non-perturbative regime. 
Let us consider first the boson component. In the absence of coupling with fermions, and for a boson gas parameter  $\eta=n_{\rm B} a_{\rm BB}^3 \ll 1$, bosons can be described at $T=0$ by Bogoliubov theory, corresponding to the values $ 8 \pi a_{\rm BB} n_0 /m_{\rm B}$   and $4\pi a_{\rm BB} n_0 /m_{\rm B}$ for the normal and anomalous self-energies, respectively (where  $n_0$ is the condensate density  and we set $\hbar=1$ throughout). 
On the other hand, previous work  for the {\em normal} phase shows that pairing correlations between bosons and fermions can be included   
rather accurately by a T-matrix type of self-energy \cite{Fra10,Fra12}.   We extend this self-energy to the condensed phase  by adding the contribution $\Sigma_{\rm BF}$ of Fig.~\ref{diagrams}(a) to the Bogoliubov contribution in the normal self-energy  $\Sigma_{\rm B}^{11}$.  The many-body T-matrix (T) appearing in  $\Sigma_{\rm BF}$ extends  to the condensed phase the corresponding T-matrix ($\Gamma$) used in the normal phase~\cite{Fra10,Fra12} by including condensate lines, as  represented by the  diagrams (c) and (d) of  Fig.~\ref{diagrams} ~\cite{footnote}. 
We neglect here any diagram containing more than one T-matrix:  pairing contributions are then excluded from the anomalous self-energy  $\Sigma_{\rm B}^{12}$. 
Feynman's rules for the finite temperature formalism then yield in the zero temperature limit:
\begin{eqnarray}
\Sigma_{\rm{B}}^{11}(\bar{k})&=&\Sigma_{\rm{BF}}(\bar{k}) + \frac{8\pi a_{ \rm{BB}}}{m_B}n_0\label{sigB11}\\
\Sigma_{\rm{B}}^{12}(\bar{k})&=&\frac{4\pi a_{ \rm{BB}}}{m_B}n_0\label{sigB12}\\
\Sigma_{\rm {BF}}(\bar{k})&=&\int\!\!\frac{d {\bf P}}{(2\pi)^{3}} \int\!\!\frac{d \Omega}{2\pi} {\rm T}(\bar{P})G_{\rm F}^{0}(\bar{P}-\bar{k}),
\end{eqnarray}
where
\begin{eqnarray}
{\rm T}(\bar{P})^{-1}&=&\Gamma(\bar{P})^{-1}-n_0G_{\rm F} ^0(\bar{P})
\label{T}\\
\label{gamma}
\Gamma(\bar{P})^{-1}&=&\frac{m_{r}}{2\pi a_{\rm BF}}-\frac{m_{r}^{\frac{3}{2}}}{\sqrt{2}\,\pi}\!\left[\frac{P^2}{2M}-2\mu -i\Omega\right]^{\frac{1}{2}}\!\!\!
-I_{\rm F}(\bar{P})\phantom{a}\\
I_{\rm F}(\bar{P})&\equiv & \int\frac{d\mathbf{p}}{(2\pi)^{3}}\,\frac{\Theta(-\xi^{\rm F}_{\mathbf{P}-\mathbf{p}})}{\xi^{\rm F}_{\mathbf{P}-\mathbf{p}}+\xi^{\rm B}_{\mathbf{p}}-i\Omega}.
\end{eqnarray}

In the above expressions we have introduced a 4-vector notation $\bar{P}\equiv(\mathbf{P},i\Omega)$,  $\bar{k}\equiv(\mathbf{k},i\omega)$, where ${\bf P}$,${\bf k}$ 
are momenta and $\Omega$,$\omega$ are frequencies. The  bare Green's functions are given by $G^{0}_{s}(\bar{k})^{-1}=i\omega-\xi^{s}_{{\bf k}}$,  where $\xi^s_{\bf p}=p^2/2m_s-\mu_s$  and $s={\rm B},{\rm F}$, while $\mu\equiv(\mu_{\rm B}+\mu_{\rm F})/2$ and $m_r=m_{\rm B}m_{\rm F}/(m_{\rm B}+m_{\rm F})$. A closed form expression for  $I_{\rm F}(\bar{P})$ is reported in \cite{Fra12}.  

The fermionic self-energy is due only to the coupling with bosons. In this case,  the T-matrix can be closed in the diagram either with a boson propagator or with two condensate insertions. The second choice, however, produces in general {\em improper} self-energy diagrams, which would lead to a double-counting when inserted in the Dyson's equation for the dressed fermion Green's function. 
{\em Proper} diagrams are obtained by replacing T with $\Gamma$ in this contribution, as shown in Fig.~\ref{diagrams}(b). The fermionic self-energy is then given by:
\begin{equation}
\label{selff}
\Sigma_{\rm F}(\bar{k})=n_0 \Gamma(\bar{k}) - \int\!\!\frac{d {\bf P}}{(2\pi)^{3}} \int\!\!\frac{d \Omega}{2\pi}  {\rm T}(\bar{P})G_{\rm B}^{0}(\bar{P}-\bar{k}).
\end{equation}
The self-energies (\ref{sigB11}), (\ref{sigB12}), and (\ref{selff}) determine the dressed boson and fermion Green's functions, once inserted in the corresponding Dyson's equations:
\begin{equation}
\label{gbog}
G'_{\rm B}(\bar{k})^{-1}= i\omega-\xi _{\mathbf{k}} ^{\rm B}-\Sigma _{ \rm B}^{11}(\bar{k})+
\frac{\Sigma_{\rm B}^{12}(\bar{k})^2}{i\omega +\xi _{\mathbf{k}} ^{\rm B}+\Sigma _{\rm B}^{11}(-\bar{k})}\phantom{aa}
\end{equation}
and
$G_{\rm F}(\bar{k})^{-1}= G^{0}_{\rm F}(\bar{k})^{-1} -\Sigma _{\rm F}(\bar{k})$.
The momentum distribution functions are in turn obtained by an integration over $\omega$: $n_{\rm F}({\bf k}) =  \int  \frac{d\omega}{2\pi} G_{\rm F}(\bar{k})\, e^{i\omega 0^+}$  and  $n_{\rm B}({\bf k})=- \int  \frac{d\omega}{2\pi} G'_{\rm B}(\bar{k})\, e^{i \omega 0^+}$
where ${\bf k}\neq 0$ for the bosons. A further integration over ${\bf k}$  yields the fermion density $n_{\rm F}=\int \!\!\frac{d \mathbf{k}}{(2\pi)^3} n_{\rm F} (\mathbf{k})$ and
the out-of-condensate density $n_{\rm B}'=\!\int\!\! \frac{d \mathbf{k}}{(2\pi)^3} n_{\rm B} (\mathbf{k})$, to which the condensate density $n_0$ must be added to get the boson  density $n_{\rm B}=n_{0} + n_{\rm B}'$. These T-matrix approximation (TMA) equations are finally supplemented by the Hugenholtz-Pines relation \cite{Hug59}:  
%\begin{equation}
%\label{hp}
$\mu_{\rm B}=\Sigma_{\rm B}^{11}(0)-\Sigma_{\rm B}^{12}(0)$,
%\end{equation}
 which, together with the above number equations, allows one to determine $\mu_{\rm B}$, $\mu_{\rm F}$, and $n_0$ for given values of $n_{\rm B}$ and $n_{\rm F}$.

 {\em Quantum Monte Carlo (QMC) method} - We estimate the momentum distributions also with the Variational Monte Carlo (VMC) and FNDMC methods, which stochastically solve the Schr\"odinger equation either with a variational wave function $\Psi_T$, or with an imaginary-time-projected wave function $\Psi_{\tau}=e^{-\tau \hat{H}}\Psi_T$, whose nodal surface is constrained to that of $\Psi_T$ so to circumvent the fermionic sign problem \cite{Reynolds1982}.  We estimate $n_s(k)$ in VMC with $n_s^{V}(k)=\langle \Psi_T|\hat{n}_k|\Psi_T\rangle/\langle \Psi_T|\Psi_T\rangle$, where $\hat{n}_k$ is the number operator in momentum space averaged over direction, while FNDMC provides the mixed estimator $n_s^{M}(k)=\lim_{\tau\to\infty}\langle \Psi_T|\hat{n}_k|\Psi_{\tau}\rangle/\langle \Psi_T|\Psi_{\tau}\rangle$. A common way to reduce the bias introduced by $\Psi_T$ in the mixed estimator is to perform the extrapolations $n_s^{E1}=(n_s^{M})^2/n_s^{V}$ or $n_s^{E2}=2n_s^{M}-n_s^{V}$, where the dependence on $\delta\Psi=\Psi_{\infty}-\Psi_T$ is second order. In practice, we use the differences  between $n_s^{E1}(k)$ and $n_s^{E2}(k)$  as a systematic error on top of the statistical error.  
 Simulations are carried out in a box of volume $L^3=N_{\rm F}/n_{\rm F}$ with periodic boundary conditions, with a number of fermions up to $N_{\rm F}=57$ and a number of bosons $N_{\rm B}$ varying with $x$. Details of the model potentials are the same as in \cite{Ber13}. We use a trial wave function  of the form $\Psi_T({\bf R})=~J({\bf R})\Phi({\bf R})$, where $J({\bf R})=\prod_{\alpha,i}f_{\rm BF}(r_{\alpha i})\prod_{\alpha,\beta}f_{\rm BB}(r_{\alpha\beta})$ is a Jastrow function of the fermionic (Latin) and bosonic (Greek) coordinates and $\Phi$ is a Slater determinant of plane waves for the fermions.
At distances $r < \bar{R}_{ss^\prime}$, the functions $f_{ss^\prime}$ are determined by solving the relevant two-body problems. For $r>\bar{R}_{BF}$, $f_{\rm BF}(r)=\exp[-u(r)-u(L-r)+2u(L/2)]$ with $u(r)=c_0+c_1/r+c_2/r^2$, where $c_0$ and $c_1$ are fixed by continuity at  $\bar{R}_{BF}$, while $\bar{R}_{BF}$ and $c_2$ are variational parameters to be optimized \cite{Casula2004}. We set $\bar{R}_{BB}=L/2$.
 \begin{figure}[t]
 \begin{center}
\includegraphics[angle=0,width=8.0cm]{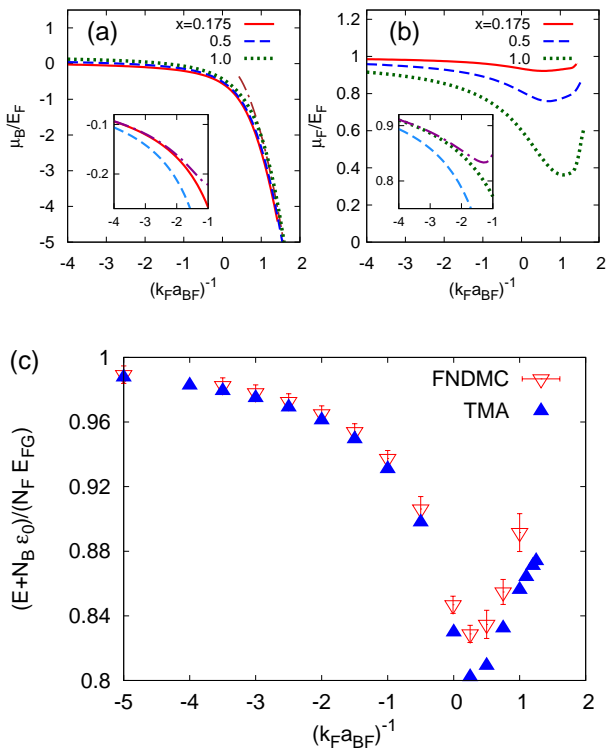}
\caption{(a) Bosonic chemical potential $\mu_{\rm B}$  vs.~$(k_{\rm F} a_{\rm BF})^{-1}$  for $m_{\rm B}=m_{\rm F}$,  $\eta= 3\times 10^{-3}$, and  different values of $x$. Dashed-dotted line: $-\epsilon_0$. (b) Fermionic chemical potential $\mu_{\rm F}$ for same parameters.
Insets: comparison at $a_{\rm BB}=0$ with 1$^{\rm st}$ (dashed) and 2$^{\rm nd}$ order (dashed-dotted) perturbative results in weak coupling for (a) $\mu_{\rm B}$ at $x=0.175$  (b) $\mu_{\rm F}$ at $x=1$.  (c)
Energy vs.~$(k_{\rm F} a_{\rm BF})^{-1}$  at $x=0.175$ and  $\eta= 3\times 10^{-3}$, with the binding energy contribution subtracted for $a_{\rm BF} >  0$.}
\label{thermo}
\end{center}
 \end{figure}

 {\em Results} -  Figure~\ref{thermo}(a) and (b) report the coupling dependence of $\mu_{\rm B}$ and $\mu_{\rm F}$ (normalized to the Fermi energy $E_{\rm F}=k_{\rm F}^2/2m_{\rm F})$ as obtained by solving the TMA equations for $m_{\rm B}=m_{\rm F}$,  $\eta= 3\times 10^{-3}$,  and three different values of $x$.   The chemical potential $\mu_{\rm B}$ tends  to 
 the mean-field value $4 \pi a_{\rm BB} n_0/m_{\rm B}$ in the weak-coupling limit $(k_{\rm F} a_{\rm BF})^{-1} \ll -1$, while it approaches $-\epsilon_0$, where $\epsilon_0=(2m_r a_{\rm BF}^2)^{-1}$ is the binding energy  of the two-body problem, when pairing correlations dominate.   
In the inset,  one can see that our calculated values of $\mu_{\rm B}$ (full line) approach the 2$^{\rm nd}$ order perturbative expression $\mu_{\rm B}=2\pi a_{\rm BF}n_{\rm F}/m_r(1+ 3 \frac{k_{\rm F} a_{\rm BF}}{2\pi})$ (dashed-dotted line) of Refs.~\cite{Alb02,Viv02}.
The fermionic chemical potential $\mu_{\rm F}$ has instead a non-monotonic behavior. For increasing attraction, it first decreases, following the 2$^{\rm nd}$ order perturbative expression $\mu_{\rm F}=E_{\rm F}+2\pi a_{\rm BF}n_{\rm B}/m_r(1+ 2 \frac{k_{\rm F} a_{\rm BF}}{\pi})$ in the weak-coupling limit (see inset),  and then increases for $(k_{\rm F} a_{\rm BF})^{-1} \gtrsim 1$, suggesting a  repulsion between unpaired fermions and correlated BF pairs, similar to that occurring in the molecular limit.

\begin{figure}[t]
\begin{center}
\includegraphics[angle=0,width=7.5cm]{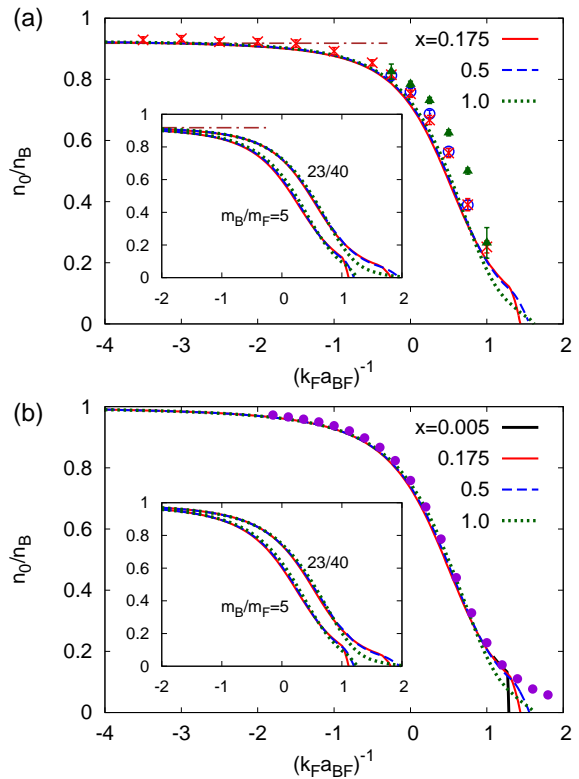}
\caption{Condensate fraction vs.~$(k_{\rm F} a_{\rm BF})^{-1}$ for different $x$.  (a)  Results for $m_{\rm B}=m_{\rm F}$,  $\eta= 3\times 10^{-3}$. Lines: TMA data and (dashed-dotted line)  Bogoliubov result $1- 8/3(\eta/\pi)^{\frac{1}{2}}$. Symbols: QMC data for $x=0.175$ (crosses), 0.5 (circles), 1 (triangles). 
(b) Results for $m_{\rm B}=m_{\rm F}$,  $a_{\rm BB}= 0$. Circles: diag-MC results of Ref.~\cite{Vli13} for $Z$. Insets: results for  $m_{\rm B}/m_{\rm F}=5$, 23/40, with same boson repulsions as in the main graphs.}
\label{condensate}
\end{center}
 \end{figure}
 
Figure~\ref{thermo}(c) compares the TMA results for the total energy $E$ (normalized to the energy of the free Fermi gas $N_{\rm F}  E_{\rm FG}$, where $E_{\rm FG}=3 E_{\rm F}/5$) with the FNDMC results for the energy in the superfluid phase for $x=0.175$ and $\eta= 3\times 10^{-3}$~\cite{Ber13} . The TMA energy  is obtained from the relation $\mu_{\rm B}= dE/dN_B$ by integrating  $\mu_{\rm B}$ from $n_{\rm B}=0$ to $n_{\rm B}=0.175 n_{\rm F}$ at fixed $n_{\rm F}$, $k_{\rm F} a_{\rm BB}$, and $k_{\rm F} a_{\rm BF}$.  
One sees that the TMA energy follows rather closely the FNDMC data (which are upper bounds to the ground-state energy) even in the fully non-perturbative regime $|k_{\rm F} a_{\rm BF}| > 1$.
Notice that to emphasize discrepancies, the binding energy contribution $-N_{\rm B} \epsilon_0$ has been subtracted to both FNDMC and TMA data for $a_{\rm BF} > 0$.  

\begin{figure}[t]
\begin{center}
\includegraphics[angle=0,width=7.5cm]{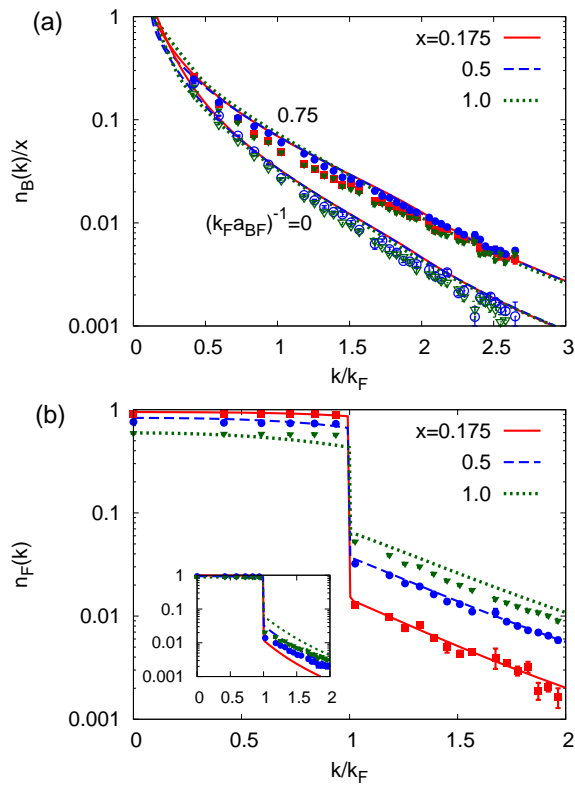}
\caption{(a) Bosonic momentum distribution function $n_{\rm B}(k)$ divided by $x$ vs.~$k$ for  $m_{\rm B}=m_{\rm F}$,  $\eta= 3\times 10^{-3}$, $(k_{\rm F} a_{\rm BF})^{-1}=0,0.75$, and different values of  $x$. Curves: TMA results. Symbols: QMC data for $x=0.175$ (crosses), 0.5 (circles), 1 (triangles). (b) $n_{\rm F}(k)$ vs.~$k$ for the same parameters as in (a) and $(k_{\rm F} a_{\rm BF})^{-1}=0.75$ (main panel) and 0 (inset).}
\label{distributions}
\end{center}
 \end{figure}

We pass now to discuss the results for the condensate fraction $n_0/n_{\rm B}$. A striking feature of Fig.~\ref{condensate}(a), reporting $n_0/n_{\rm B}$ vs.~$(k_{\rm F}a_{\rm BF})^{-1}$ for different $x$ and constant $\eta$,  is that  the curves calculated within  TMA  at different concentrations 
collapse on top of each other for most of their graph (specifically, deviations from this universal behavior occur for  $n_0/n_{\rm B}\lesssim 0.2$ where, however, the condensed phase is no longer the ground state, according to the phase diagram of~\cite{Ber13}).  This occurs not only for $m_{\rm B}= m_{\rm F}$, but also for different mass ratios (the inset reports examples  for $m_{\rm B}/m_{\rm F}= 5, 23/40$, the latter value corresponding to a $^{23}$Na-$^{40}$K mixture). 
Our QMC simulations confirm this universality for $ x \le 0.5$, with results very close to TMA. Deviations appear instead for $x=1$,  with larger values of $n_0/n_{\rm B}$ compared to the results at lower concentrations (or to TMA), with the exception of the point at ($k_{\rm F}a_{\rm BF})^{-1}=1$, which has however large error bars due to uncertainties in the QMC extrapolation method at this or larger couplings. Part of this discrepancy could be ascribed to the lack of information on molecular correlations in the nodal surface of $\Psi_T$, with a consequent increase of $n_0/n_{\rm B}$
due to an underestimate of the pairing effects, especially at high concentration where interaction effects on the fermions are more important. Moreover, finite-size effects and the use of Jastrow wave functions generally increase $n_0$ of QMC calculations \cite{Reatto1969}, which we thus consider as an upper bound.

 The universality of the condensate fraction just found with both methods for $x\le0.5$  prompts us to consider the limit $x\to 0$, and  establish a connection with the problem of a single impurity immersed in a Fermi sea (the `polaron problem'  that much attention has  received recently in the context of polarized Fermi gases \cite{Pro08,Vei08,Mas08,Mor09,Pun09,Com09,Sch09,Mat11,Koh12,Vli13}).
 What is the analogous of the condensate fraction for the polaron problem? 
 %We argue that the quasi-particle weight $Z$ of the polaron coincides with the limiting value of the condensate fraction of a BF mixture for $x\to 0$. 
 Consider first the polaron as the $x\to 0$ limit in a BF mixture. By definition $n_0/n_{\rm B} = n_{\rm B}(k=0)/N_{\rm B}$, then reducing to $n_{\rm imp}(k=0)$ for $x\to 0$  (where $n_{\rm imp}(k)$ is the momentum distribution of a single impurity). Regard now the polaron as the high polarization limit  of an imbalanced  Fermi gas, and focus on the  
  quasiparticle weight $Z$ at the Fermi  momentum  $k_{{\rm F} \downarrow}$ of the minority component ($\downarrow$).
  The weight $Z$ determines the height of the Fermi step:
 $Z=n_{\downarrow}(k_{{\rm F} \downarrow}^-)-n_{\downarrow}(k_{{\rm F} \downarrow}^+)$. 
For vanishingly small concentration $k_{{\rm F} \downarrow} \to 0$ and $n_{\downarrow}(k)\to n_{\rm imp}(k)$, then yielding
$Z=n_{\rm imp}(k=0)-n_{\rm imp}(0^+)=n_{\rm imp}(k=0)$ for $V\to \infty$. This is because $n_{\rm imp}(k\neq 0)$  scales like $V^{-1}$, since its integral scales like the density of one particle in the volume  $V$.  We thus conclude that  for $x\to 0$ the condensate fraction of a BF mixture tends to the polaron quasiparticle weight $Z$. Figure~\ref{condensate}(b) compares then our data for the condensate fraction at different $x$ (and $\eta=0$ as for the polaron problem)  with the  diagrammatic Monte Carlo data for the polaron quasiparticle weight $Z$ reported in~\cite{Vli13}. We see that the curve at the lowest concentration follows indeed the data for $Z$ for all couplings, until it vanishes almost with a jump at a critical coupling (indicating a real jump for $x=0^+$). In addition, due to the universality discussed above, also the curves at larger concentrations follow the polaron weight $Z$, with deviations just in their ending part, where they vanish more gently than at low concentrations. Note further, by comparing Fig.~3(a) and (b), that in the coupling region $(k_{\rm F} a_{\rm BF})^{-1} \ge 0$ of most interest, the boson repulsion has a minor effect on $n_0/n_{\rm B}$.
By measuring the condensate fraction in a BF mixture, even at sizable boson concentrations, one would thus obtain $Z$ in a completely different and independent way from the radio-frequency spectroscopy or Rabi oscillations techniques used for imbalanced Fermi mixtures~\cite{Sch09,Koh12}.     

The universal behavior of $n_0/n_{\rm B}$ suggests to look for a similar behavior in the whole $n_{\rm B}(k)$.  To this end, we divide $n_{\rm B}(k)$ by the concentration $x$, as shown in Fig.~\ref{distributions}(a) for both TMA and  QMC calculations.  The results obtained by the two methods agree well and show that curves and data obtained at different concentrations almost collapse on top of each other.  
For the fermionic momentum distributions $n_{\rm F}(k)$ of
 Fig.~\ref{distributions}(b), the agreement  between QMC and TMA results is slightly worse.  This can be attributed to finite size effects, which are more severe for the fermionic momentum distributions (see the detailed discussion of these effects of Ref.~\cite{Hol11}).
   
 In conclusion, we have presented a diagrammatic approach for the condensed phase of a BF mixture which compares well with QMC calculations over an extended range of boson-fermion couplings, including the fully non-perturbative region $|k_{\rm F} a_{\rm BF}| > 1$.  By using both methods, we have found that the condensate fraction and the bosonic momentum distributions are ruled by curves  which, in an extended concentration range,  are universal with respect to the boson concentration. We have also found an unexpected connection between the condensate fraction in a BF mixture and the quasiparticle weight of the Fermi polaron, unifying in this way features of polarized Fermi gases and BF mixtures.  
 
 \acknowledgements
We acknowledge CINECA and Regione Lombardia, under the LISA initiative, for the availability of high performance computing resources and support. Financial support from the University of Camerino under the project FAR ``CESN"  is also acknowledged.


\begin{thebibliography}{99}
\bibitem{footnote}
A simpler T-matrix approach for the condensed phase  could be constructed by using $\Gamma$ in the place of T in the self-energies (3) and (7). This approach is however correct only to first order in the weak-coupling limit of the BF interaction, while our T-matrix approach is consistent to second order.   We have  also explicitly verified that this alternative T-matrix approach yields quite different (and rather unphysical) results for the condensate fraction compared to Monte Carlo calculations and the present T-matrix approach.
\end{thebibliography}
\end{document}